# A Nuclear Equation of State Inferred from Stellar *r*-process Abundances

Erika M. Holmbeck[1,2,3,5], Richard O'Shaughnessy[1], Vera Delfavero[1], and Krzysztof Belczynski[4]
[1] Center for Computational Relativity and Gravitation, Rochester Institute of Technology, Rochester, NY 14623, USA; eholmbeck@carnegiescience.edu
[2] Observatories of the Carnegie Institution for Science, 813 Santa Barbara St., Pasadena, CA 91101, USA
[3] Joint Institute for Nuclear Astrophysics—Center for the Evolution of the Elements (JINA-CEE), USA
[4] Nicolaus Copernicus Astronomical Center, Polish Academy of Sciences, ul. Bartycka 18, 00-716 Warsaw, Poland
*Received 2021 October 11; revised 2021 December 15; accepted 2022 January 4; published 2022 February 24*

## Abstract

Binary neutron star mergers (NSMs) have been confirmed as one source of the heaviest observable elements made by the rapid neutron-capture (*r*-) process. However, modeling NSM outflows—from the total ejecta masses to their elemental yields—depends on the unknown nuclear equation of state (EOS) that governs neutron star structure. In this work, we derive a phenomenological EOS by assuming that NSMs are the dominant sources of the heavy element material in metal-poor stars with *r*-process abundance patterns. We start with a population synthesis model to obtain a population of merging neutron star binaries and calculate their EOS-dependent elemental yields. Under the assumption that these mergers were responsible for the majority of *r*-process elements in the metal-poor stars, we find parameters representing the EOS for which the theoretical NSM yields reproduce the derived abundances from observations of metal-poor stars. For our proof-of-concept assumptions, we find an EOS that is slightly softer than, but still in agreement with, current constraints, e.g., by the Neutron Star Interior Composition Explorer, with $R_{1.4} = 12.25 \pm 0.03$ km and $M_{\mathrm{TOV}} = 2.17 \pm 0.03\, M_\odot$ (statistical uncertainties, neglecting modeling systematics).

*Unified Astronomy Thesaurus concepts:* Nucleosynthesis (1131); R-process (1324); Population II stars (1284); Neutron stars (1108)

## 1. Introduction

Metal-poor stars in the Galaxy, deficient in iron produced by supernovae (SNe), provide records of the nucleosynthetic events that first enriched the interstellar medium and the prenatal gas out of which these stars formed. One of these nucleosynthetic signatures found among very metal-poor stars ($[\mathrm{Fe/H}] < -2.0$)[6] is the elemental production by rapid neutron capture (the *r*-process). Stars with extreme levels of enrichment by the *r*-process account for only about 5% of metal-poor stars in the Milky Way's halo (Barklem et al. 2005; Beers & Christlieb 2005). Due to their low iron content, the source of the relatively high abundances of trans-iron elements implies a high yield, potentially short delay time type of nucleosynthetic event (Cowan et al. 1991; Argast et al. 2004; Cowan et al. 2021, and references therein).

The observational confirmation of *r*-process production by neutron star mergers (NSMs) through GW170817 (Abbott et al. 2017) and its corresponding lightcurve SSS17a/AT 2017gfo (Coulter et al. 2017; Cowperthwaite et al. 2017; Drout et al. 2017; Kilpatrick et al. 2017; Shappee et al. 2017) demonstrates at least one astrophysical site for the source of *r*-process elements in the universe (Kasen et al. 2017). Questions still remain, however, about whether double neutron star (DNS) systems can merge with sufficiently short delay times and if they can produce sufficient *r*-process yields to account not only for the majority of the abundances in individual metal-poor stars, but also for the abundance *spread* among metal-poor stars with *r*-process elements, e.g., in [Eu/Fe] (Matteucci et al. 2014; Cescutti et al. 2015; Ishimaru et al. 2015; Côté et al. 2019; Safarzadeh et al. 2019).

Some evidence, albeit indirect, for a prolific *r*-process site like NSMs as the progenitors of the *r*-process rich metal-poor stars can be found outside the Galaxy. The discovery of the large fraction of stars with extreme *r*-process enrichment in the ultrafaint dwarf galaxy Reticulum II that far exceeds the Milky Way's own mere 5% point to small satellite galaxies as the natal birth sites of metal-poor stars (Ji et al. 2016; Roederer et al. 2016). Stars with appreciable levels of *r*-process enrichment, but not quite as high as those in Reticulum II, could have originated in more massive ultrafaint galaxies, as evidenced by the identification of such a star in Tucana III (Hansen et al. 2017). The Galactic halo's metal-poor stars having originated in small, early galaxies is also commensurate with the hierarchical merger origin of the Milky Way (Searle & Zinn 1978; Schlaufman et al. 2009; Tumlinson 2010). In these small galaxies, a prolific (perhaps single) event like an NSM could eject enough *r*-process material to account for the high [Eu/Fe] abundances observed in systems like Reticulum II and Tucana III (e.g., Safarzadeh & Scannapieco 2017; Safarzadeh et al. 2019; but see also Tarumi et al. 2020). Due to their low mass, such systems would have low star formation rates and could retain their metal-poor nature for longer, possibly alleviating the necessity that NSMs must have short delay times.

Although NSMs are so far the only *r*-process site with direct observational confirmation, contributions by black hole–NS mergers (NSBHs), by exotic types of SNe (collapsars, or magneto-hydrodynamic jet-driven SNe), and other theoretical sites (e.g., dark matter induced NS implosions; Bramante & Linden 2016; Fuller et al. 2017) should still be quantified in order to understand the evolution of elements in the Galaxy (see, e.g., Cameron 2003; Pruet et al. 2004; Surman & McLaughlin 2004; Winteler et al. 2012; Nishimura et al. 2015; Mösta et al. 2018; Siegel et al. 2019; Miller et al. 2020). In addition, even assuming NSMs are the only r-process formation sites, many incompletely understood mechanisms are critical ingredients in any forward model to predict the r-process abundance in the universe, not only in metal-poor stars, but also in our solar system. For example, the production of the heavy elements observed in stars fundamentally relies on the microphysics of nuclei far from stability, the magneto-hydrodynamics and (neutrino) radiative transfer of merging compact objects, and an estimate of how often merging NS binaries form over all cosmic time. A comprehensive answer to this complex problem requires a full integration of theory with observation and experiment, both in astrophysics and in nuclear physics. In this work, we focus on the microphysics in the production of elements in NSMs: how the nuclear equation of state (EOS) shapes NSM yields.

---

[5] Hubble Fellow.
[6] $[\mathrm{A/B}] = \log(N_A/N_B)_* - \log(N_A/N_B)_\odot$, where $N$ is the number density of an element in the star (*) compared to the Sun (⊙).

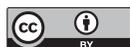

Instead of a forward-modeling approach by predicting expected r-process yields from theory, recent work has used observed r-process abundances to constrain nuclear physics inputs. Such "reverse engineering" has been accomplished for, e.g., predicting the masses of rare-earth nuclei from the solar r-process abundances, assuming a variety of thermodynamic conditions (Mumpower et al. 2017; Vassh et al. 2021), and by solving for the masses of the progenitor merging NSs from stellar abundance measurements (Holmbeck et al. 2021a). In this paper, we take a similar reverse-engineering approach to build an entirely phenomenological EOS inferred by the r-process abundances in metal-poor stars. Section 2 describes the population of metal-poor stars and their r-process abundances from which we will reverse engineer a neutron star EOS. Next, the theoretical input for our model framework is discussed in Section 3, including the population of coalescing NS binaries, descriptions of their ejecta yields, the model for the EOS itself, and constraints that we will apply on the EOS. Our reverse-engineering approach and how all of the theoretical and observational aspects work together to build an EOS is described in Section 4, and Section 5 presents the results from our reverse-engineering model.

## 2. Observations of Metal-poor Stars

The elemental r-process pattern is surprisingly robust for neutron-rich conditions (such as that found in NSMs; e.g., Korobkin et al. 2012; Just et al. 2015), with two significant exceptions: the lightest r-process elements—$_{38}$Sr, $_{39}$Y, and $_{40}$Zr—and the heaviest—$_{90}$Th and $_{92}$U. The near-constancy of the r-process pattern is observationally supported by the r-process enhanced, metal-poor stars themselves (Cowan et al. 1999; Sneden et al. 2008). When scaled, the abundance patterns of these stars are in nearly perfect agreement among the lanthanide elements, but up to 1 dex of variation can exist at the extrema of the r-process pattern, i.e., at Sr and Th (Mashonkina et al. 2014; Siqueira Mello et al. 2014; Ji et al. 2016). These elements can be extremely sensitive to r-process conditions like the initial electron fraction, $Y_e = n_n(n_p + n_n)^{-1}$, of the ejecta (Eichler et al. 2019; Holmbeck et al. 2019, 2021a). Therefore, we seek a sample of metal-poor stars with measurements of Sr and Th to quantify the extent of composition variation in r-process enhanced metal-poor stars. We also choose stars with Dy to compare the Sr and Th production to the lanthanide abundance of the r-process patterns. Unfortunately, definitive measurements of only 29 stars with these three elements exist currently in the literature. These 29 stars are the same as in Holmbeck et al. (2021a) and can be found in Table 2 therein. Requiring Th inherently biases our sample toward stars that are already highly r-process enhanced ([Eu/Fe] > +0.7), since these high-enhancement stars only constitute a minority of metal-poor stars with r-process elements. If we seek to account for the r-process abundances in the *majority* of metal-poor stars, we need to consider the much larger population of metal-poor stars that do not have Th measurements.

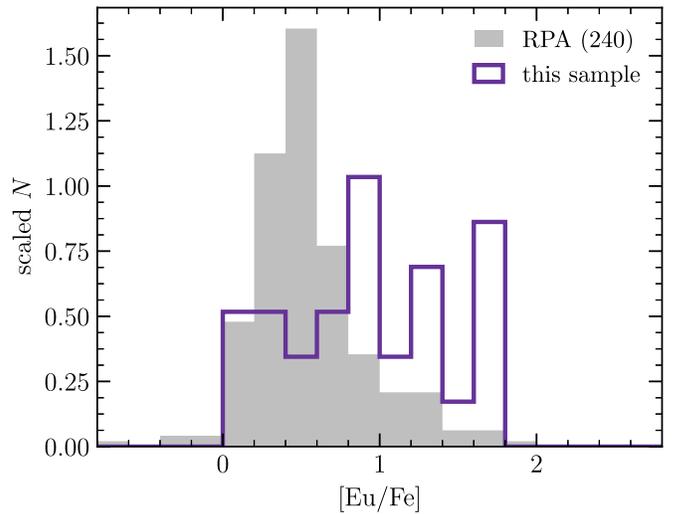

**Figure 1.** Normalized [Eu/Fe] histograms of stars with measurements of Sr, Dy, and Th (purple line) compared to a larger sample of metal-poor stars.

The R-Process Alliance (RPA) has released Sr, Ba, Eu, and Fe abundance determinations for nearly 600 stars based on "snapshot" high-resolution spectra. These abundances are sufficient to quantify the extent of r-process enhancement within each star. First, we take stars in all RPA data releases to date (Hansen et al. 2018; Roederer et al. 2018; Sakari et al. 2018a, 2018b, 2019; Ezzeddine et al. 2020; Holmbeck et al. 2020) and take those that lie within a similar metallicity range as the initial 29-star sample ($-3.3 \lesssim$ [Fe/H] $\lesssim -1.5$). From this trimmed list, we then pick stars that have [Ba/Eu] < $-0.5$ to eliminate those that could have obtained their heavy elements from the *slow* neutron-capture process rather than the r-process. These cuts leave 240 stars with definitive Ba and Eu measurements (i.e., no upper/lower limits) from the RPA spanning a wide range of [Eu/Fe] abundances. We take this sample as representative of the frequency of various levels of r-process enrichment in the Galaxy. Figure 1 shows our 29 star sample compared to these 240 r-process stars in the RPA. The 29 stars with Th measurements skew toward higher values of [Eu/Fe], while the stars in the RPA sample on average favor lower [Eu/Fe] values. We interpret this skew as observational in nature; stars already enhanced in r-process elements will have correspondingly high actinide abundances, allowing Th to be detected more readily in their spectra. Therefore, we assume that Th ought to be present in the remaining 240 RPA stars as well. Ideally, we would use these 240 stars directly, but since most do not have reported Th measurements, we adjust the 29 star sample to occupy a similar [Eu/Fe] distribution as the much larger RPA sample.

First, we bin the data in [Eu/Fe] abundance bins of 0.2 dex and count how many stars in the RPA and 29 star samples are in each bin. The 0.2 dex bin size is on the order of the typical [Eu/Fe] abundance uncertainty (see, e.g., Holmbeck et al. 2020). The ratio of the abundance histograms of the 29 star sample and 240 star sample in [Eu/Fe] abundance provide a weight by which to scale the 29 star sample and shape it to be representative of a broader range of metal-poor stars with $r$-process elements that do not presently have Th measurements. This rescaling inherently assumes that, although the 29 star sample is not representative in [Eu/Fe], it is representative in [Th/Eu]. Next, we build the observational sample based on these weights.

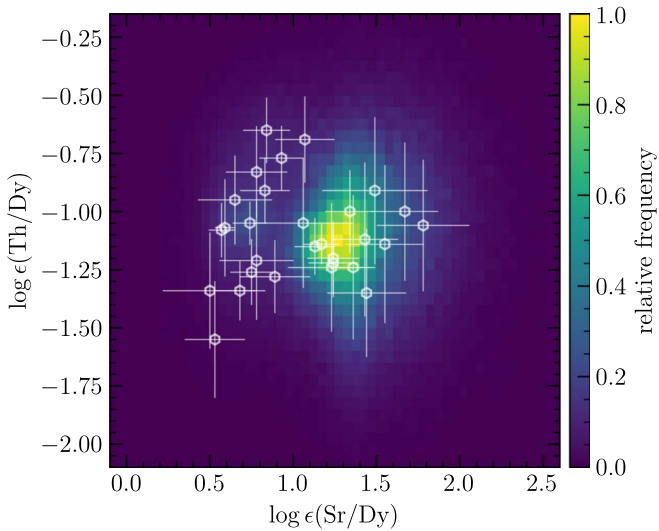

**Figure 2.** Observational values of $\log \epsilon$ (Sr/Dy) and $\log \epsilon$ (Th/Dy) from 29 stars (white points) and the inferred distribution computed by the rescaling process (colors).

To build our "complete" observational sample, we place the 29 stars in $\log \epsilon$ (Sr/Dy) versus $\log \epsilon$ (Th/Dy) abundance space,[7] accounting for measurement error and the rescaling weights described above. Our estimate for the density is therefore a weighted two-dimensional kernel density estimate: $p(\mathbf{a}) \propto \sum_i w_i K(\mathbf{a} - \mathbf{a}_i, \Sigma)$, where $K(x, \Sigma)$ is a normal distribution, $\Sigma$ is the diagonal covariance matrix reflecting abundance errors (typically about 0.2 dex), and $w_i$ are the weights identified via the ratios above. We do not propagate statistical uncertainty in the scaling weights. In practice, we build this density estimate via a weighted Monte Carlo procedure. For each of the 29 stars, we randomly scatter $N$ points drawn from two Gaussians, representing the two abundance ratios. The centroid and $1\sigma$ spread of the Gaussian correspond to the star's abundance measurement and reported $1\sigma$ uncertainty (typically less than 0.2 dex). Then, the two-dimensional Gaussian is scaled by the weight found previously from the ratio of the [Eu/Fe] bin counts. Figure 2 shows the combined two-dimensional histogram after this series of random selections and rescaling. The original 29 stars are also shown. Note that, although many stars with Th measurements have comparatively low Sr/Dy abundances ($\lesssim 1.0$), the high [Eu/Fe] ratios in these stars are not very common, as shown in Figure 1. Therefore, the population of resampled observations with low Sr is diminished, and the relatively higher Sr/Dy abundance signature occurs more frequently.

---

[7] Here, $\log \epsilon (A) = \log(N_A/N_H) + 12$, where $N_A$ is the number density of element $A$.

This resampled population is an attempt to describe the expected Sr/Dy and Th/Dy abundances of metal-poor stars in the Galaxy that do not presently have measurements for these elements. We will use this population as the "observations" from which to build an EOS, essentially requiring that the EOS effect on NSM ejecta is such that the total NSM yields reproduce this two-dimensional distribution.

## 3. Theoretical Model

We can construct a forward model for the $r$-process abundances in metal-poor stars from some assumptions about how and how many NSs merge (a DNS population) and the physical properties of those NSs (determined by the EOS). To summarize and visualize how the masses and EOS enter the elemental yield calculation, Figure 3 shows a schematic of the input, output, and intermediate steps that relate NS masses and an EOS to abundance observations in our model. First, a DNS population and EOS are chosen, with the EOS being the free parameter. The three different cases we explore in this work (discussed in the following subsections) differ in the input DNS distribution. Then, the NS masses and the EOS enter into a series of analytic functions and nucleosynthesis network models to finally output total elemental yields for all NSMs in the DNS population list. Finally, these model output abundances are compared to the observational abundances discussed in Section 2 to determine how successful the chosen EOS is at reproducing the observations. Using a Markov Chain Monte Carlo (MCMC) algorithm, a new EOS is chosen and this process repeated to generate a posterior on the unknown EOS for each (three) of our input DNS variations. We expand on each of these steps in the following sections.

### 3.1. DNS Populations

This section describes the DNS populations we use for this study, which will combine with the EOS input to determine specific NSM yields. First, we use a theoretical formation model (population synthesis) to estimate the past history of NSMs in our Galaxy. We will apply this DNS model to two cases in this work. Then, we consider the case for the current DNS population, with the underlying assumption that the $r$-process producing NSMs were of systems similar to present-day DNS systems in the Milky Way.

The continuously updated StarTrack code estimates the evolution of single and interacting binary stars, both individually and as populations (Belczynski et al. 2008). Frequently applied to interpret (and calibrated against) astronomical observations of several types—including binary pulsars and gravitational-wave observations—this code provides a useful benchmark to explore plausible self-consistent models for the Galactic binary population. After reviewing several recent studies (Dominik et al. 2012, 2013, 2015; Belczynski et al. 2016a, 2020, 2016b; Wysocki et al. 2018; Drozda et al. 2020), we choose the M15 model (submodel B) in Belczynski et al. (2016b) with strong pair-instability (pulsation) SNe and modest NS natal kicks ($\sigma = 130$ km s$^{-1}$; Wysocki et al. 2018). This model successfully reproduces compact object merger rates from the third observing run of the Laser Interferometer Gravitational-Wave Observatory (LIGO) and Virgo collaboration, remains qualitatively consistent with known constraints on the maximum NS mass ($M_{\rm TOV}$), and is consistent with (most) observed Galactic pulsar masses.

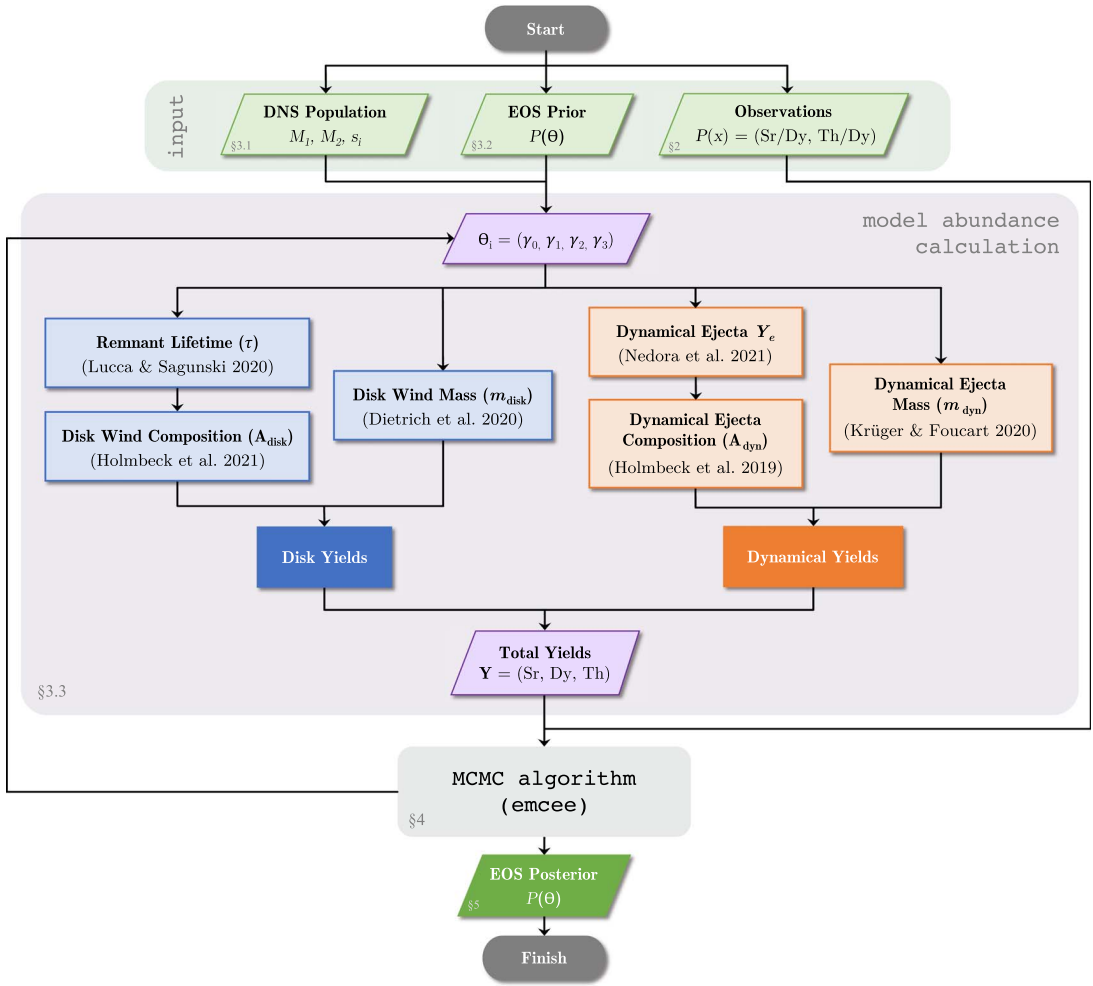

**Figure 3.** Schematic flow of the model calculations and an indication in which section each is discussed.

We translate the M15 simulation (of a fiducial cosmological volume, as in Dominik et al. 2015; Belczynski et al. 2016a) to our application of the Milky Way as follows. Since we are interested only in the *distribution* and not the (highly uncertain) overall scale factor, we disregard overall normalization. In addition, because DNSs typically merge fairly quickly, we can ignore small differences between cosmological and galactic star formation history. We can therefore identify the relative likelihood of NSMs by the weighted data ($M_{1,i}$, $M_{2,i}$, $s_i$), where $s_i$ represents the star formation, metallicity, and population-weighted likelihood of the $M_{1,i}$–$M_{2,i}$ mass pair, as described in previous work by Dominik et al. (2015). For computational efficiency, we keep only those mass pairs that are in the top 99% of the most frequent merging systems in the population synthesis model, i.e., those with $s_i \geqslant 0.01 \times \max(\vec{s_i})$. Of several thousand unique mass combinations, these top 99% account for roughly 50% of all unique merging DNS systems in the population synthesis model output. Although we are potentially ignoring cases that might eject significant amounts of *r*-process material, our choice to work in abundance *ratios* rather than absolute abundances effectively removes the sensitivity to total ejecta mass. At most, we are removing cases that would contribute at the 1% level in Sr/Dy−Th/Dy abundance space, at the benefit of doubling our computational efficiency.

Figure 4 shows the masses of the merging DNSs in the StarTrack model in purple, weighted by their likelihood ($s_i$). The most likely merging system has $M_1 \approx 1.35\ M_\odot$ and $M_2 \approx 1.1\ M_\odot$. Note that there are no systems in this model with $M_{1,2} > 2.0\ M_\odot$, and none within the first mass gap, as a consequence of the model choices in Belczynski et al. (2016b) and Wysocki et al. (2018). We will test two cases using this DNS population: one with a constraint on the maximum mass of a nonrotating NS ("with $M_{\rm TOV}$") and one without such a constraint ("no $M_{\rm TOV}$"). These two cases will be discussed in more detail in Section 3.2.

Next, instead of using a synthetic DNS population, we can also use the observed mass distribution of NSs in the Milky Way. As a third, alternative, case we reconstruct the EOS informed by inferences about the mass distribution of individual Galactic NSs. Specifically, we assume the primary NS mass is drawn from a mixture of two normal distributions, with a primary peak at $1.34\ M_\odot$ and a secondary at $1.78\ M_\odot$ (from Alsing et al. 2018; see also Özel et al. 2012). We draw the secondary mass based on the mass-ratio distribution from Farrow et al. (2019): essentially a Gaussian peaked at $q = 1$, where $q > 0.69$ with 99% confidence. We choose 500 random selections from these distributions to represent the primary and secondary NS masses in the binary system. This DNS population is equivalent to assuming that the NSM progenitors of the *r*-process material in the Galaxy were of DNS systems similar to existing DNS binaries currently known in the Milky Way. The primary and secondary masses for this empirical distribution are also shown in Figure 4 in yellow. On average, this hypothesized Galactic DNS distribution predicts a much broader total mass distribution for DNSs than the observed distribution of known Galactic DNSs (see, e.g., Farrow et al. 2019), but extends to total masses more consistent with GW190425 (Abbott et al. 2020). Except for the input DNS population, this third case ("Alsing+") will use the same method and constraints as the "with $M_{\rm TOV}$" case that uses the M15 model for the DNS masses.

Considering that the metal-poor stars likely originated in ultrafaint dwarf galaxies like Reticulum II and Tucana III, the correlation between [Fe/H] and time becomes unclear and prevents imposing a delay-time distribution on these merger pairs. Rather, we instead consider the cumulative effect that the entire population of NSMs would have had on the r-process abundance ratios.

### 3.2. EOS Parameters

Within the framework of this paper, the EOS bridges the DNS masses with theoretical descriptions of NSM yields and, therefore, with a total abundance distribution that can be compared to observations of metal-poor stars. Since the EOS is the unknown parameter, we will use an MCMC method to build an EOS from the abundance observations themselves. First, let us discuss a basis for building an EOS. To smoothly explore an EOS parameter space, we consider a four-parameter spectral-decomposition representation, described by Lindblom (2010) and implemented in LALSUITE (LIGO Scientific Collaboration 2018). The energy density as a function of the pressure, $\epsilon(p)$, is described by the adiabatic index, $\Gamma$:

$$\Gamma(p) = \frac{\epsilon + p}{p}\frac{dp}{d\epsilon}. \qquad (1)$$

In the spectral-decomposition representation, the EOS can be expressed as an expansion of the adiabatic index:

$$\Gamma(x) = \exp\left(\sum_k \gamma_k x^k\right), \qquad (2)$$

where $x = \log(p/p_0)$, and $p_0$ is the minimum pressure. In the LALSUITE implementation, $k$ goes from 0 to 3, and the EOS is therefore defined by four $\gamma_k$ values. With some constraints on these parameters, arbitrary EOSs and NS mass–radius curves can be constructed. LALSUITE provides a robust set of built-in functions to construct a neutron star EOS from $\vec{\gamma}$ and find properties such as the radius and tidal deformability as functions of the NS masses. Our method utilizes this capability of LALSUITE to smoothly explore the $\vec{\gamma}$ parameter space and find the EOS-dependent properties needed by the ejecta yield descriptions.

There are several constraints from both dense-matter theory and NS observations available to guide the EOS parameter choices. In addition, not all allowed combinations of $\vec{\gamma}$ lead to physical EOSs. First, rather than consider the entire $\vec{\gamma}$ parameter range nominally allowed by LALSUITE, we use the transformation and corresponding limits from Appendix B of Wysocki et al. (2020) to explore only those combinations that are physically allowed. These transformed parameters will be referred to as $\vec{\gamma}'$.

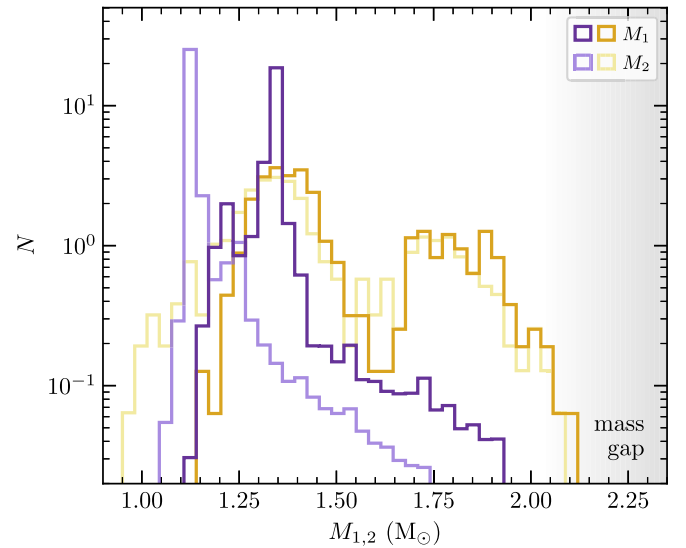

**Figure 4.** Masses of the primary and secondary NSs in the M15 StarTrack model (purple) and the Galactic distribution (gold).

Next, we fold in constraints from observation and theory. There are no direct constraints on the $\vec{\gamma}$ parameters since these are merely the representations to build the EOS (but see Jiang et al. 2020). However, solving the TOV equations with the EOS parameters does allow direct comparison with astronomical observables. Of primary importance is the maximum nonrotating NS mass, $M_{\rm TOV}$. We use the updated mass of PSR J0740+6620 as a lower limit that $M_{\rm TOV}$ must satisfy for the EOS (Fonseca et al. 2021; Riley et al. 2021). This limit is implemented by adopting a cumulative distribution function (CDF) of a normal distribution ($\mathcal{N}$) with the centroid and $1\sigma$ spread from the measurements of PSR J0740+6620, namely, CDF($x$, $\mathcal{N}(\mu, \sigma)$), where $\mu = 2.072$ and $\sigma = 0.07$. This CDF enforces $M_{\rm TOV}$ of at least $2.072\,M_\odot$—with some room for uncertainty at the lower end—but imposes no upper limit. An upper limit on $M_{\rm TOV}$ has been suggested from the remnant mass of the GW170817 merger event based on inferences that the remnant probably collapsed somewhat quickly ($\lesssim 0.1$ s) into a black hole (e.g., Margalit & Metzger 2017; Shibata et al. 2019; Kawaguchi et al. 2021). One of our model cases enforces $M_{\rm TOV}$ at a somewhat higher $2.30 \pm 0.15\,M_\odot$ limit ("with $M_{\rm TOV}$") to include this observational constraint. As a second case ("no $M_{\rm TOV}$"), we release the upper limit on $M_{\rm TOV}$ and allow the model to explore solutions with arbitrarily high NS masses, accommodating the possibility that the GW170817 remnant did survive longer.

On the microphysics side, both theory and experiment offer some constraints on the EOS. These constraints place limits on, e.g., the symmetry pressure, $L$, and the nuclear incompressibility, $K_\infty$, at saturation densities ($\rho_0 \approx 2.8 \times 10^{17}$ kg m$^{-3}$, or $n_0 \approx 0.17$ fm$^{-3}$). Tight constraints on $K_\infty$ exist from experiment; however, these values are defined for symmetric nuclear matter, whereas the EOSs defined with LALSUITE assume neutron star material in $\beta$-equilibrium. Therefore, we implement limits on the somewhat more loosely constrained, but more analytically straightforward, parameter $L$. For an EOS, the pressure can be expressed as

$$p(n) = n^2 \frac{d\epsilon}{dn}, \qquad (3)$$

where $n$ is the nucleon number density instead of the mass density, $\rho$. On a plot of energy density per nucleon versus number density, $L$ is proportional to the slope at nuclear saturation:

$$L = 3n_0 \frac{d\epsilon}{dn}\bigg|_{n_0} \quad (4)$$

$$= \frac{3}{n_0} p(n_0). \quad (5)$$

In this form, $L$ can be straightforwardly evaluated with the functions available in LALSUITE. Energy density versus density curves can be calculated directly by LALSUITE; however, we caution that these values are the *volumetric* energy and *mass* densities, respectively. Therefore, we take the pressure-versus-mass densities from LALSUITE and convert the mass density to a number density by dividing by the nucleon mass plus the typical binding energy per nucleon ($\approx -16$ MeV). Note that this approximation is only valid around nuclear saturation. In addition, this transformation to obtain $L$ may be different from the "true" value by a few MeV. In this way, we can obtain $L$ for an EOS using the functions available in LALSUITE. When evaluating the likelihood of a particular EOS (discussed in Section 4), we calculate $L$ from Equation (5) and prefer EOSs within a liberal range of 30 MeV $\lesssim L \lesssim$ 100 MeV, as determined by theory and experiment (e.g., Dutra et al. 2012; Lattimer 2012; Tews et al. 2019; Piekarewicz 2021). These limits are imposed as a likelihood in the model, which will be discussed in Section 4.

### 3.3. Ejecta Yields

With a population of merging DNSs in hand (two cases to be tested with the StarTrack M15 model and one with the Alsing et al. 2018 distribution), we can compute the total, EOS-dependent $r$-process yields for each merger in our DNS list. We use Equation (6) from Krüger & Foucart (2020) to quantify the NSM dynamical ejecta and Equation (S4) from Dietrich et al. (2020) for the disk mass. The amount of mass that is lost through winds (neutrino-driven or viscous) from the accretion disk is assumed to be a constant fraction of 40% of the total disk mass as in, e.g., Radice et al. (2018). (We will discuss the implications of this assumption in Section 5.) In addition to depending on the NS masses, these fit choices also depend directly on physical properties of NSs such as the radius, compactness, and tidal deformability. Since these properties are EOS-dependent, the ejecta fits can be expressed as functions of the NS masses and the EOS, so the net ejecta can be expressed as $m_{\rm ej}(M_1, M_2, \vec{\gamma}) = m_{\rm dyn} + 0.4\, m_{\rm disk}$, where $\vec{\gamma}$ represents the EOS parameters, described in Section 3.2.

The available fits to NSM ejecta come with significant uncertainties, mostly as a result of fitting across multiple groups and simulations that may differ in their implementations. To account for these uncertainties, we allow the computed ejecta masses and lifetimes to vary from (truncated) normal distributions. Each distribution is centered at the expected value given by each equation and has a spread corresponding to the uncertainty of the fit (for details, see Radice et al. 2018; Krüger & Foucart 2020; Lucca & Sagunski 2020). For each DNS merger pair in the population synthesis output, we draw $N = 200$ samples from this truncated-normal distribution such that each NSM gives some statistical spread of masses rather than one discrete value; i.e., in the expressions above, each $m_{\rm dyn}$ and $m_{\rm disk}$ is a set of length $N$ that accounts for uncertainty in each corresponding relationship. We then multiply each of these ejecta masses by the appropriate trajectory-dependent nucleosynthesis output to produce overall outflow yields. We also allow for a similar spread in the corresponding nucleosynthetic compositions produced by the dynamical and post-merger outflows to obtain a distribution of outflow yields. All nucleosynthesis output is based on previous calculations in Holmbeck et al. (2019) and Holmbeck et al. (2021a) that use the FRDM2012 nuclear mass model (Möller et al. 2012) for theoretically calculated nuclear reaction and decay rates.

For the dynamical ejecta composition, we use the thermodynamic evolution of a trajectory from the 1.4–1.4 $M_\odot$ NSM simulations of S. Rosswog (Piran et al. 2013; Rosswog et al. 2013) as in Korobkin et al. (2012). The initial composition may vary based on the EOS and masses of the merging NSs, so we choose final abundances from a nucleosynthesis network calculation that starts with a merger-dependent $Y_e$ as in Holmbeck et al. (2019), where the initial $Y_e$ is calculated using Equation (10) from Nedora et al. (2021). Since a spread in $Y_e$ may be expected in the dynamical ejecta, we apply some variation on the initial $Y_e$. The Gaussian is centered at the calculated $Y_e$ and given a 1$\sigma$ spread of 0.01, motivated by the values in Table 1 of Nedora et al. (2021). Therefore, we calculate the final abundances for a range of initial $Y_e$ values and combine those abundances with weights from a normal distribution. We draw the same number of samples ($N = 200$) as for the ejecta masses from this normal distribution to obtain a set of abundances, $A_{\rm dyn}$, of dimension $N \times 3$, where there are $N$ yields for each of the three elements we study here: Sr, Dy, and Th.

Similarly, the composition of the post-merger outflows may be influenced by the lifetime ($\tau$) of the remnant massive NS before it collapses into a BH (Lippuner et al. 2017). We draw random samples of the lifetime of the merger remnant based on the uncertainty in the lifetime fit from Lucca & Sagunski (2020). Then, lifetime-dependent $r$-process yields from Holmbeck et al. (2021a) are used to find the ultimate merger-dependent disk wind abundances, $A_{\rm disk}$.

Finally, these random samples drawn from $Y_e$ and $\tau$ for the dynamical and disk compositions are multiplied with their corresponding ejecta masses and added to obtain the final yields:

$$Y = (A_{\rm dyn} m_{\rm dyn}) + 0.4(A_{\rm disk} m_{\rm disk}).$$

Note that $Y$ also has dimensions of $N \times 3$. For each merging DNS pair, the Sr/Dy and Th/Dy abundance ratios are computed by an element-wise division of the corresponding columns in $Y$. The final $N$-sized set of two abundance ratios is added to a two-dimensional distribution in abundance space as in Figure 2, after being weighted by the likelihood factor, $s_i$, for that merging pair. (For the Alsing+ case, $s_i = 1$.) Including these systematic uncertainties of the ejecta masses, the remnant lifetime, and the dynamical ejecta $Y_e$ significantly increases computation time, but it allows flexibility in the model that would otherwise be too highly constrained by the literature fits to the simulations, especially as those fits update with new available data.

## 4. MCMC Method

Above we have connected theoretical merger outflows to stellar *r*-process abundance observations through a four-parameter, spectral-decomposition EOS. In this section, we describe how we will find the EOS parameters ($\vec{\gamma}$) that, given a particular DNS population (M15 and the Alsing et al. 2018 distribution), best match observed *r*-process abundances in metal-poor stars. We use an MCMC method to sample these parameters and find the spectral-decomposition representation that would best reproduce the abundance distribution. In our calculation, we only reconstruct an abundance *distribution*, not the overall amount of *r*-process material. Therefore, all scale factors associated with uncertainties in the total amount of material ejected and in the DNS merger rate factor out of our conclusions.

We build an EOS from $\vec{\gamma}'$ as described in Section 3.2 and choose a prior on $\vec{\gamma}'$ such that $R_{1.4}$ is uniform. In terms of this coordinate system, our likelihood has four factors: a lower limit on $M_{\rm TOV}$ from pulsar observations; an optional upper limit on $M_{\rm TOV}$ from GW170817; a constraint on the symmetry energy, $L$; and, finally, an estimate of the likelihood of current observations, given our model. The first three terms in the likelihood are represented as a product of cumulative distribution functions:

$$\mathcal{L}_{\rm obs} \equiv {\rm CDF}_{\rm PSR}(1 - {\rm CDF}_{M_{\rm TOV}}){\rm CDF}_L. \quad (6)$$

The first CDF represents the applied constraints on the lower limit of $M_{\rm TOV}$ from the mass measurement of PSR J0740 +6620:

$${\rm CDF}_{\rm PSR} = {\rm CDF}(M_{\rm TOV}, \mathcal{N}(2.072, 0.07)).$$

The second term corresponds to an optional upper limit on $M_{\rm TOV}$, covering the cases we explore here.

$${\rm CDF}_{M_{\rm TOV}} = \begin{cases} {\rm CDF}(M_{\rm TOV}, \mathcal{N}(2.3, 0.15)) & \text{with } M_{\rm TOV} \\ 0 & \text{no } M_{\rm TOV} \\ {\rm CDF}(M_{\rm TOV}, \mathcal{N}(2.3, 0.15)) & \text{Alsing+} \end{cases}$$

Lastly, the constraint on $L$ is given a very broad CDF from theory:

$${\rm CDF}_L = {\rm CDF}(L, \mathcal{N}(30, 5))[1 - {\rm CDF}(L, \mathcal{N}(100, 5))].$$

We use CDFs instead of a hard limit so that the parameter space still has a finite probability of exploring even extreme regions.

Because we have already performed a nonparametric estimate of the abundance distribution from the underlying observations, we cannot compare our forward model to each individual observation with single-event log-likelihoods (e.g., $\ln p(x_k)$). Instead, we start with the KL divergence (Kullback & Leibler 1951) to estimate the population-averaged difference in log-likelihood when drawing a fixed number of events from this sample. The average per observation increment in log-likelihood will be (see, e.g., O'Shaughnessy 2013)

$$\langle \ln \mathcal{L} \rangle = D_{\rm KL}(P|Q),$$

where $P$ is the observations, $Q$ is the model, and $D_{\rm KL}$ is the KL divergence between the two:

$$D_{\rm KL}(P|Q) = \int dx\, P(x) \ln\left(\frac{P(x)}{Q(x)}\right).$$

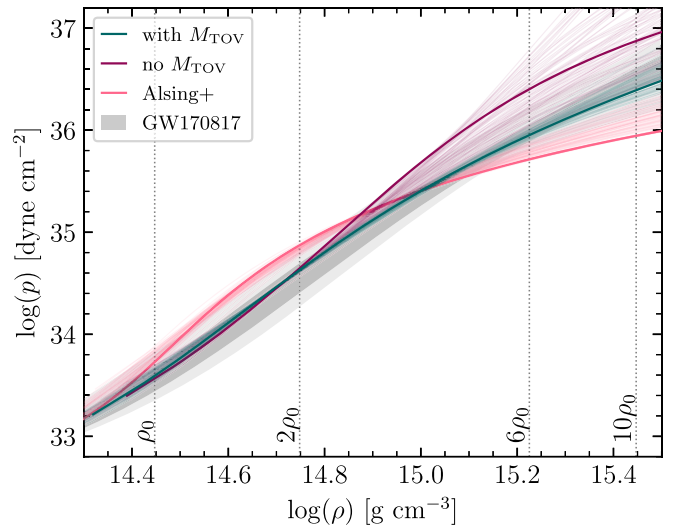

**Figure 5.** Posterior samples (thin lines) and maximum likelihood (thick lines) NS EOSs for the case in which a constraint on $M_{\rm TOV}$ is used (teal) and when there is no such constraint (pink). The gray band indicates the 50% (darker) and 90% (lighter) confidence intervals for the EOS inferred from GW170817.

To account for the effective number of observations entering into the divergence from the original RPA sample, we can multiply this log-likelihood by an overall constant. Because this expression will go to infinity if the two distributions do not have matching support, we for practical applications instead use the Jensen−Shannon (JS) divergence (Lin 1991):

$$D_{\rm JS} = \frac{1}{2}D_{\rm KL}(P|M) + \frac{1}{2}D_{\rm KL}(Q|M),$$

where $M = (P + Q)/2$. Thus, for the likelihood of *r*-process abundances, we use

$$\mathcal{L}_r \equiv e^{-n_{\rm eff} D_{\rm JS}}, \quad (7)$$

where we adopt $n_{\rm eff} = 20$. The total likelihood when we include other constraints on the EOS is therefore $\mathcal{L} = \mathcal{L}_{\rm obs}\mathcal{L}_r$:

$$\mathcal{L} = {\rm CDF}_{\rm PSR}(1 - {\rm CDF}_{M_{\rm TOV}}){\rm CDF}_L\, e^{-n_{\rm eff} D_{\rm JS}}. \quad (8)$$

All that is left now is to run the MCMC sampler to explore the $\vec{\gamma}'$ parameter space. For 32 walkers, a burn-in stage of 500 steps is more than sufficient for convergence. After burn-in, we allow the MCMC sampler to continue for another 3000 steps, resulting in nearly 100,000 individual samples for our three cases: the M15 model for the DNS population (1) with and (2) without a constraint on $M_{\rm TOV}$, and (3) the Alsing et al. (2018) DNS distribution with the $M_{\rm TOV}$ constraint.

## 5. Results and Discussion

Figures 5 and 6 show the EOS and mass–radius curves, respectively, for the MCMC posteriors of our three cases using the functions available to LALSUITE. The maximum-likelihood solutions for each case are shown as thick lines, while 100 random samples from the posterior distributions are shown as thin, faint lines. The median and standard deviation of the $M_{\rm TOV}$, $R_{1.4}$, and $L$ for each of the three posteriors are listed in Table 1. The entire posterior distributions for the three cases can be found at Holmbeck et al. (2021b). In Figure 5, the inferred EOS from GW170817 is shown in gray (Abbott et al. 2019). The contours in Figure 6 show posterior distributions from NICER and XMM-Newton measurements (PSR J0030+0451 from Miller et al. 2019 and PSR J0740+6620 from Miller et al. 2021). Neither constraints on the EOS from GW170817 nor on the NS radius from the NICER/XMM-Newton measurement were included in our model.

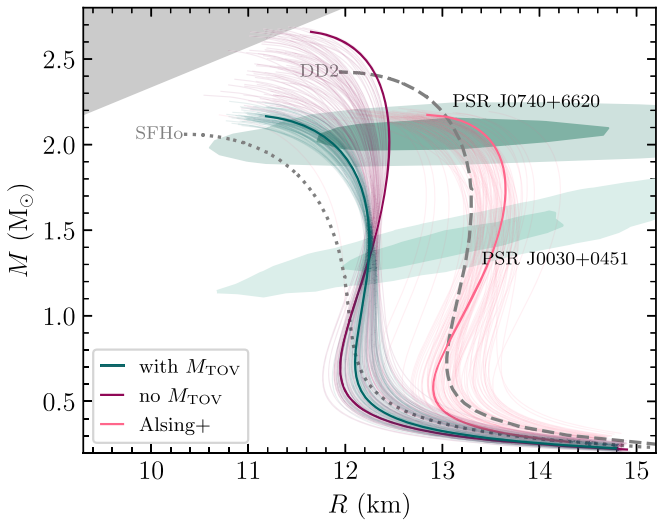

**Figure 6.** Posterior samples (thin lines) and maximum likelihood (thick lines) mass–radius curves for the case in which a constraint on $M_{\rm TOV}$ is used (teal) and when there is no such constraint (pink). For comparison, two existing theoretical EOSs (SFHO, dotted, and DD2, dashed) are shown, along with recent pulsar measurements from NICER (shaded contours). The upper-left gray shaded region indicates where causality is violated ($R > 2.9\ GM$).

**Table 1**
Median and One-sigma Confidence Interval for Neutron Star and EOS Properties for Each Case

| Model | $M_{\rm TOV}$ ($M_\odot$) | $R_{1.4}$ (km) | $L$ (MeV) |
| --- | --- | --- | --- |
| with $M_{\rm TOV}$ | $2.17 \pm 0.03$ | $12.25 \pm 0.03$ | $44 \pm 4$ |
| no $M_{\rm TOV}$ | $2.46 \pm 0.15$ | $12.28 \pm 0.04$ | $46 \pm 5$ |
| Alsing+ | $2.18 \pm 0.02$ | $13.47 \pm 0.41$ | $64 \pm 9$ |

### 5.1. M15: with $M_{TOV}$ and no $M_{TOV}$

For the two cases that use the M15 model for the DNS population, Figure 5 shows some agreement with the EOS derived for GW170817. The two sets of posterior samples are tightly constrained at low densities (perhaps due to the constraint on $L$), but diverge at higher densities. This divergence is unsurprising since NSs are reasonable probes of the low-density EOS, and only the most massive NSs offer insight into densities greater than $6\rho_0$ (Lattimer 2012).

The mass–radius curves derived for these EOSs in Figure 6 are slightly softer than the NICER/XMM-Newton median value, though there is still reasonable agreement within $2\sigma$ of their derived radii. Including agreement with the NICER/XMM-Newton in the likelihood function does not significantly change these results. Both M15 cases lead to EOS solutions narrowly grouped in mass–radius at $M \sim 1.35\ M_\odot$, most likely do the high frequency of DNSs with $M_1 = 1.35\ M_\odot$ in the M15 model (see Figure 4). However, for the "no $M_{\rm TOV}$" case, the posterior solutions sometimes violate causality, as indicated by the pink curves extending into the gray region in Figure 6. The two families of solutions are nearly identical in the ejecta parameters they predict, indicating some degeneracy when the spectral-decomposition parameters are propagated to ejecta and abundance observables through the NS mass–radius relationships. For the most common case in the StarTrack model of a 1.35–1.1 $M_\odot$ merger, both cases predict a wind mass of about $9 \times 10^{-2}\ M_\odot$ and a much smaller dynamical mass of about $6.5 \times 10^{-3}\ M_\odot$. The two cases also give the same dynamical ejecta $Y_e$ of about 0.17. Therefore, a 1.35–1.10 $M_\odot$ merger would likely produce the same kilonova signature with either of the two maximum-likelihood solutions in the with-/no-$M_{\rm TOV}$ cases. For a 1.40–1.35 $M_\odot$ merger, the predicted wind mass differs by about 20% between the two solutions, which could leave a distinguishable signature in the associated kilonova.

Figure 7 shows the Th/Dy and Sr/Dy abundances and the difference between the model and observations produced by the maximum-likelihood solution for the "with $M_{\rm TOV}$" case. The model reasonably reproduces the most prominent feature in the abundance space of Figure 2 occurring at Sr/Dy $\approx 1.25$ and Th/Dy $\approx -1.15$, though the peak Th/Dy is somewhat lower. A couple of distinguishable artifacts also appear in the output abundances: first, a "spur" at high Sr/Dy. This feature commonly occurs for quite low-mass, low-asymmetry mergers. Because of their low total mass, the wind outflows from the long-lived remnant accretion disk drive the Sr abundances to high values. That said, the dynamical masses for mergers with $M_1 \sim 1.2\ M_\odot$ and $q \gtrsim 0.9$ will be small, but not zero. Although the dynamical mass may be very low, actinide production can still be sufficiently high to contribute a significant fraction of the total Th that is ejected. This spur therefore represents a limiting case in which low asymmetries and low masses drive not only high Sr production, but also a low-mass tidal tail that is rich in actinides.

Second, with the ejecta equations employed here, there appears to be a diagonal floor in which the ejecta cannot simultaneously be rich in Sr/Dy and deficient in Th/Dy. Conditions producing this lower edge are ones with $q \gtrsim 0.9$ and total masses $M_1 + M_2 \approx 2.6$–2.9. At the very lowest Th/Dy are mergers with $M_{1,2} \sim 1.45\ M_\odot$. Conditions close to the solar abundances are typically produced by mergers with $M_{1,2} \sim 1.3\ M_\odot$. These conditions produce long-lived remnants that have the same wind compositions and relatively small wind ejecta masses. Going from the upper right of the edge to the lower left (i.e., increasing the total binary mass), the contribution by the dynamical ejecta is increased. In nearly symmetric cases, low tidal deformabilities lead to dynamical $\langle Y_e \rangle$ values that struggle to produce significant amounts of Th, but have a high yield of Dy. Their low-Th and high-Dy abundances on average bring the total Th/Dy and Sr/Dy yields down from pure-wind abundances as the NS masses increase. Therefore, this edge can be thought of as the limiting wind composition case with an increasing amount of moderately low-$Y_e$ dynamical ejecta contributing to the total outflows. Both the high-Sr/Dy spur and the low-Th/Dy floor indicate limits in the computational method from both the compositions and the total ejecta masses. However, it is worth noting that the extension of the "observations" into the low-Th/Dy, high-Sr/Dy region— below the diagonal floor—are not populated by an actual observational measurement in a metal-poor star. Rather, recall the distribution was given a random spread to account for observations that statistically *could* have these particular abundance combinations in our attempt to build a complete sample by removing some observational bias. Therefore, it may be possible that no such combination exists in metal-poor stars that have primarily *r*-process origins of their heavy element material. More observations of metal-poor stars (currently underway, e.g., through RPA efforts) could shed light on the reality of this limit.

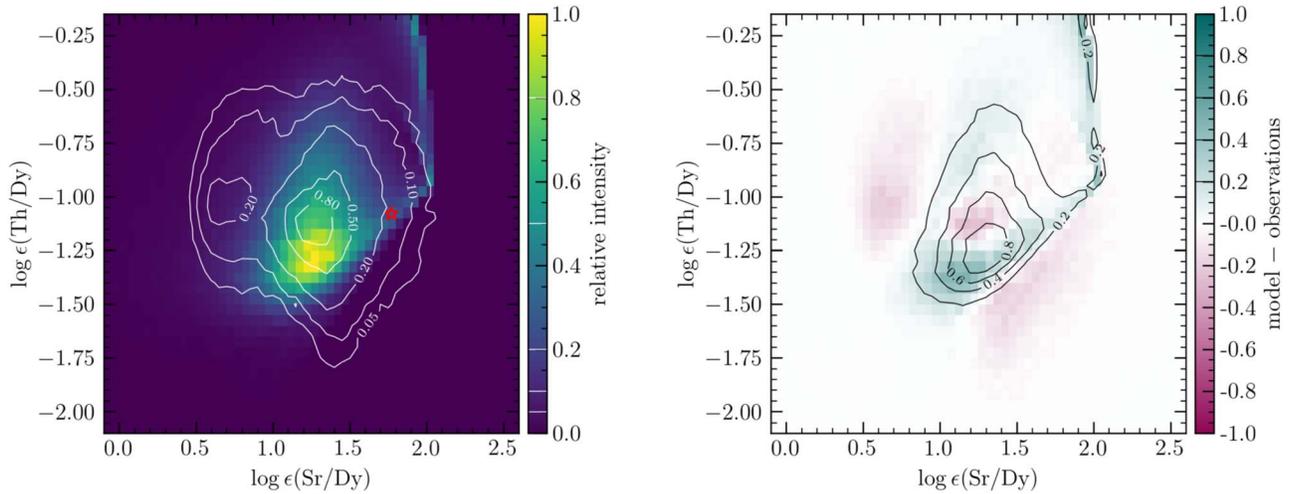

**Figure 7.** Left: model output with observational contours from Figure 2 overlaid in white. The solar value is indicated by the red star. Right: difference from observations for the maximum-likelihood model as well as normalized contours corresponding to the JSD values between the model and observations.

The abundances show some sensitivity to the percentage of the total disk mass that escapes from the system (assumed to be 40% here). Since most of the Sr that is produced in our model is from the wind component, altering this disk-to-wind fraction has the largest effect on the Sr/Dy abundances. Increasing (decreasing) this fraction effectively moves the output abundances to the right (left) on Figure 7, up to the diagonal abundance floor. In order to compensate for the surplus (deficiency) of Sr from increasing (decreasing) the ejecta fraction, the EOS must lead the merger to eject relatively less (more) Sr compared to Dy. Changing the Sr/Dy ratio can be achieved most notably by changing the NS compactness. Soft EOSs with relatively smaller NS radii will eject more dynamical ejecta for $q \gtrsim 0.9$, moderately low-mass systems, such as those in the DNS populations that use the M15 population synthesis model. With more Dy, the large Sr/Dy values that are achieved by higher disk-to-wind fractions are effectively reduced. Therefore, we expect that increasing (decreasing) the disk-to-wind fraction will lead to inferred EOSs that are more (less) compact. However, exploring the sensitivity of our model to this fraction would require a separate study, and for now we recognize it as an area for improvement.

There is one feature that is virtually absent in the model output: the lobe at moderately high Th and low Sr. To produce low Sr, the dynamical ejecta must be very high. However, NS–NS mergers with massive dynamical ejecta require high asymmetries and masses. Such massive, asymmetric cases will correspondingly produce a high Sr abundance from the disk outflows as well. In other words, in this model, a massive dynamical ejecta component will always come with a similarly massive (or even more massive) wind component. Therefore, this low-Sr/Dy region could possibly be populated if there were very little to no wind ejecta at all. An r-process source that may realize this case—and one that this model neglects—is an NSBH merger. Theoretically, disk wind outflows would not exist in the NSBH merger case, and the mass outflows would be entirely dynamical: perhaps rich in Th and poor in lighter r-process elements like Sr. Therefore, it is possible that the most Sr-poor abundances that could not be reproduced by the NS−NS mergers in this work could have NSBH origins instead. The StarTrack output also provides these NSBH binaries, and we reserve computing and including their nucleosynthetic outflows for future work.

### 5.2. Alsing+

The two cases that use the M15 model for the DNS population both produce reasonable agreement with all available data: the EOS from GW170817, the NS radius and mass measurements from NICER/XMM-Newton, and the r-process abundances of metal-poor stars. Here, we look at how the (Alsing et al. 2018) DNS distribution fares with our MCMC model to see if we can obtain additional agreement with the current distribution of DNS systems in the Galaxy.

While the two M15 cases show reasonable behavior for their posterior EOS samples, the Alsing+ case rather proceeds to very extreme values for the EOS parameters, which can be seen by the large fluctuations in the slope of the EOS in Figure 5. To reconcile the Alsing et al. (2018) DNS distribution with the observed abundances, a very extreme EOS is needed, perhaps even beyond the limits and constraints supplied to the model. For this case, there is no agreement with the GW170817 EOS at low densities. On the other hand, the posterior mass–radius curves for the Alsing+ case agree rather well with the NICER/XMM-Newton values, preferring a stiffer EOS than the cases that use the M15 model. Finally, Figure 8 shows the resulting abundances for the best-fit EOS case using the Alsing et al. (2018) NS mass distribution. There is nearly no agreement with the r-process abundance data, and the results group tightly around high Th/Dy and Sr/Dy values. This particular feature consists of the symmetric mergers with $M_{1,2} \approx 1.4$, the main peak in the Alsing et al. (2018) distribution in Figure 4. The "wisps" of abundances around the data result from the slightly asymmetric cases that appear in the Alsing et al. (2018) distribution.

The clear failure of this model indicates that this estimate of the present-day NS mass distribution (and the associated assumption that DNSs can form with primaries drawn from this distribution) cannot be reconciled with the r-process abundances in metal-poor stars, under all of the other constraints and assumptions supplied to the model. This conclusion is in contrast to earlier work (Holmbeck et al. 2021a), which claims agreement between *r*-process stars and the Galactic DNS distribution (e.g., for the DD2 EOS). The largest source of difference between that work and the present model might possibly be the description of the dynamical ejecta $Y_e$. In this model, we employ a relationship between the $Y_e$ of the dynamical ejecta and the NS properties, including the EOS (i.e., from Equation (10) of Nedora et al. 2021). Since the results are sensitive to this description, a robust and systematic study of dynamical ejecta $Y_e$ as a function of NS properties would greatly benefit this type of work.

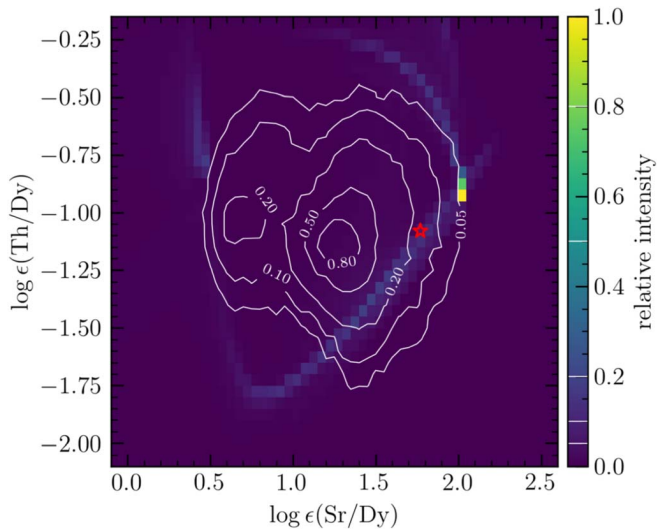

**Figure 8.** Model output for the case in which our hypothesized Galactic DNS population is used. The observational contours from Figure 2 are overlaid in white, and the solar value is indicated by the red star.

## 6. Summary and Outlook

In this work, we perform a proof-of-concept calculation showing how to combine information from metal-poor stars, population synthesis models, EOS calculations, and NSM ejecta properties to build a unique EOS for neutron stars. Our calculation estimates and propagates many recognized systematic uncertainties into our final result. Our recommended best model is the "with $M_{\rm TOV}$" case, which suggests a somewhat soft EOS. This phenomenological EOS is one in which the majority of *r*-process material in metal-poor stars can be described by mergers between binary NSs from a stellar population that simultaneously produces binary BH, NS−NS, and NS−BH mergers that agree with their inferred rates from LIGO−Virgo detections. Our EOS additionally finds agreement with measurements by NICER and EOS constraints inferred from GW170817.

As demonstrated by the failure of the "Alsing+" case to reproduce the observations, the model is quite sensitive to our input choices and does not by design guarantee a plausible outcome for the EOS. For example, using observationally inappropriate (uniform) weights for the RPA stars also leads to an implausible EOS. The agreement that our "with $M_{\rm TOV}$" model finds with NS observations, *r*-process abundances of metal-poor stars, and EOS constraints from theory and observations is a result of our specific input choices, notably the abundance distribution of metal-poor stars and the population of merging DNSs.

We recognize that several uncertainties accompany our method, notably: uncertain predictions of the mass distribution of merging NSs and whether they can populate the mass gap (Drozda et al. 2020), the dependence of the dynamical ejecta composition ($Y_e$) on the merging NS masses and EOS (including the effect of neutrino interactions; Goriely et al. 2015; Martin et al. 2018; Vincent et al. 2020), the fraction of the NSM accretion disk that can eventually become unbound, and the nuclear mass model that determines properties of *r*-process nuclei far from stability (Mumpower et al. 2016). Much work remains to be done in refining the model input for this method, and we are hopeful that recent and upcoming progress in the observational, theoretical, and experimental sectors will help uncover the fundamental structure of NSs and the properties of unstable nuclei that participate in the *r*-process. Degeneracies in our results might be broken in the future by combined effects on direct observables such as kilonovae (e.g., Barnes et al. 2021; Zhu et al. 2021). These improvements will be necessary to assess and propagate more modeling systematics into our overall uncertainty budget for the EOS.

The fact that our model results fare so well despite the model uncertainties and assumptions indicates the potential power that lies in the observed *r*-process abundances of metal-poor stars. We encourage using these stars as an additional, albeit indirect, source of data for microphysics studies of the *r*-process.

E.M.H. and R.O.S. acknowledge support from National Science Foundation (NSF) award AST 1909534. E.M.H. is additionally supported by NASA through the Hubble Fellowship grant HST-HF2-51481.001. Part of this work was performed at Aspen Center for Physics, which is supported by NSF grant PHY-1607611 and in part by a grant from the Simons Foundation. V.D. is supported by NSF grant PHY-2012057. K.B. acknowledges support from the Polish National Science Center grant Maestro (2018/30/A/ST9/00050).

*Software:* corner (Foreman-Mackey 2016), emcee (Foreman-Mackey et al. 2013), LALSuite (LIGO Scientific Collaboration 2018), Matplotlib (Hunter 2007), NumPy (van der Walt et al. 2011), SciPy (Virtanen et al. 2020).

## ORCID iDs

Erika M. Holmbeck https://orcid.org/0000-0002-5463-6800
Richard O'Shaughnessy https://orcid.org/0000-0001-5832-8517
Vera Delfavero https://orcid.org/0000-0001-7099-765X